\documentclass[prl,aps,twocolumn]{revtex4}
\usepackage{epsfig}
\begin{document}

\title{Information encoding in homoclinic chaotic systems}
\author{I.P. Mari\~no$^{*}$, E. Allaria, R. Meucci, S. Boccaletti and F.T.
Arecchi$^{\dagger}$}

\affiliation{Istituto Nazionale di Ottica Applicata, I50125 Florence-Italy} 
\affiliation{$^{*}$Also with the  
 Departamento de Ciencias Experimentales e Ingenier\'{\i}a, Universidad Rey
Juan Carlos, M\'ostoles, Madrid, Spain}
\affiliation{$^{\dagger}$Also with the Departement of Physics, University of Firenze, 
Italy} 

\date{29 August 2001}

\begin{abstract}
In this work we describe a simple  
method  of encoding information at real time in the inter-spike
intervals of a homoclinic chaotic system. This
has been experimentally tested by means of an instantaneous  synchronization
between the laser intensity of a CO$_2$ laser with feedback in the regime
of Sil'nikov chaos and an external
pulsed signal of very low power. The information is previously
encoded in the temporal intervals between
consecutive pulses of the external signal. The value of the inter-pulse
intervals is varied each time a new pulse is generated.


\end{abstract}

\maketitle

Chaotic carriers are exploited to encode and transmit signals 
via two distinct approaches. The first method
\cite{grebo1} consists in  
applying control techniques \cite{control} in order
to stabilize one of the uncountably infinite chaotic orbits embedded within the
chaotic attractor, such that its intersections with the Poincar\'e section
can be mapped into a desired sequence of bits \cite{grebo2}.
The second method, that has been more extensively used, consists in exploiting 
the synchronization
properties of coupled chaotic systems \cite{synchrony} in order to
modulate a chaotic carrier with a message signal at the transmitter, 
and then recover the message by demodulating the synchronized state
at the receiver. In the past, the latter method has been mainly focussed
on the problem of warranting privacy in the
communication \cite{various}, but it has recently found applications for
communication with chaotic time-delayed optical systems \cite{roy}.

In this Letter we discuss an alternative procedure which makes use of pulse
synchronization to encode desired messages into the interspike
interval sequences of a homoclinic chaotic system. This way, the information
is coded only in the time intervals at which spikes occur, and does not
affect any geometrical property of the chaotic flow, thus resulting in a
better performance against unwanted perturbations, 
as e.g. noise contamination in the communication
channel. We will discuss the performance of the proposed method in an
experiment on a $CO_2$ laser displaying a regime of Sil'nikov chaos.

Sil'nikov chaos \cite{shil65-70} has been observed in many systems,
such as chemical \cite{argo87} and laser \cite{arec87} experiments. This 
kind of behavior shows striking similarities with the electrical spike trains
travelling on the axons of animal neurons \cite{hodg52,feud00}. More generally,
chemical oscillators based on an activator-inhibitor competition, which
rule biological clocks controlling living rhythms, such as the heart pacemaker,
hormone production, metabolism, etc., display rhythmical trains of spikes
with erratic repetition frequency as shown in Refs. \cite{argo87,arec87}.

Homoclinic chaos of the Sil'nikov type normally appears when a parameter
is varied towards the homoclinic condition associated with a saddle focus
\cite{wigg84,kuzn95,shil65-70}. Its peculiarity consists of an astonishing
regularity of the geometric trajectory in phase space. The chaotic motion
is characterized by the large  fluctuations in the return time  of consecutive 
spikes, so that only an average return period can be defined. Thus, an 
appropriate indicator of chaos may be the distribution of the return times
to a given threshold, and the strength of chaos is associated with the
amount of decorrelation between successive returns \cite{arec88}. 
Such decorrelation occurs around the saddle focus, when the system
displays a large susceptibility, i.e. a large response to an 
external perturbation \cite{arec01}; we will
exploit this susceptibility for information encoding.

Recently, different phase locking regimes between  an initial
homoclinic chaotic signal  and a sinusoidal periodic 
external modulation has been reported \cite{alla01}. The initial homoclinic chaotic behavior can
be suppressed 
by the action of a periodic external forcing, giving rise to different phase
locking domains, (1:1, 1:2, 1:3 and 2:1), depending on the frequency of the applied 
periodic forcing.  The applied forcing readjusts the evolution of the dynamics
of the homoclinic chaotic signal in such a way that the return period of every 
orbit in phase space is the same. Thus, a periodic orbit  is stabilized.

Based on this  synchronization, we propose
an innovative and  viable method of controlling instantaneously the temporal
interval between spikes. In
this way,  the return period of each cycle changes accordingly to an external
information. In order to get an exact control
in the return period of each orbit, it is  fundamental to minimize
the response time of the system  to the forcing signal. For this purpose,
the external perturbation is made of short pulses.
In this way,  synchronization between both signals is obtained almost  
instantaneously, thus achieving  a robust control
on the individual inter-spike interval of a homoclinic chaotic system,
and hence encoding the desired information in the sequence of
inter-spikes.

\begin{figure}[!h]
\begin{center}
\leavevmode
\epsfxsize 80 mm \epsfbox{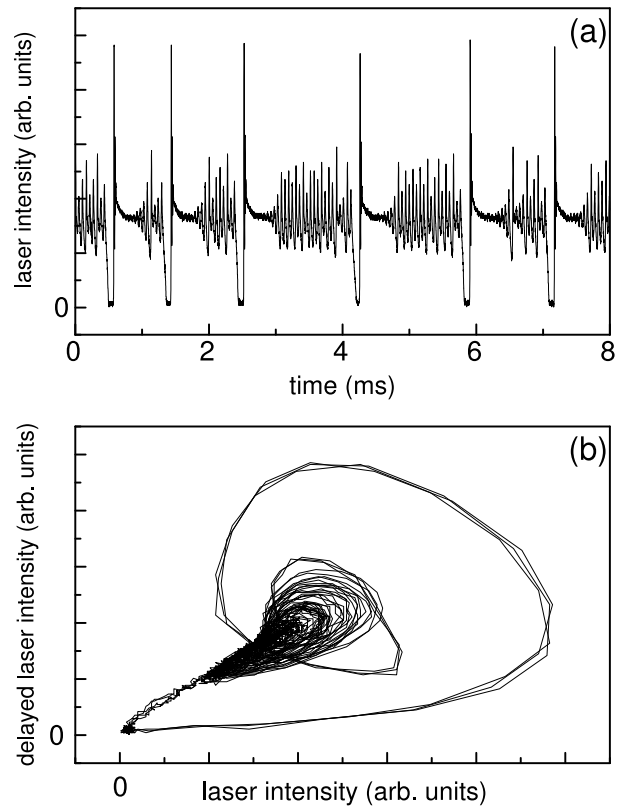}
\end{center}
\caption{(a) Experimental time series of the laser intensity for a CO$_2$ 
laser with feedback, in the regime of homoclinic chaos. The intensity modulation has been
amplified and lifted for visual convenience. 
Homoclinic chaotic behavior in a CO$_2$ laser. (b) Phase space trajectory built by an embedding technique
with appropriate delay.}
\label{fig1}
\end{figure}

The experiment has been performed on a single mode CO$_2$ laser with a
feedback proportional to the output intensity (see Ref. \cite{alla01}). The control parameters,
bias voltage and gain of the feedback, are set in a regime of homoclinic chaotic
behavior (see Fig. \ref{fig1}(a)). Figure \ref{fig1}(b) shows the phase space
reconstruction of the chaotic attractor.
The modulating external signal is applied, in an additive way, on the bias parameter with 
a waveform generator Tektronix TM5003,
which is controlled by a real time PC board (PCI-7030/6040E) from National Instruments.
The perturbation 
signal is a train of  square waveforms. The inter-pulse intervals can be controlled,
in real time, by means of an adequate computer program.  Each time a new pulse is generated, this
distance in time is changed, to encode in these time intervals the desired information.
 The inter-pulse interval is varied inside the
domain of 1:1 phase locking for the external perturbation and the laser intensity.
This  time variation range corresponds to a  frequency of  repetition of
pulses that can take values in a quite large interval.
The time duration of each pulse is $10\%$ of the inter-pulse interval. The amplitude of the pulses  is
$4\%$ of the value of the amplitude of the main spikes of the temporal
evolution of the laser intensity. The laser output is recorded by a digital
oscilloscope  with a sampling time of $1$ $\mu s$.

The first performance index we have studied  is the correlation between
the different time (or frequency) intervals of consecutive pulses of the external
signal  and  the corresponding return periods (or frequencies) in 
the temporal series of the laser intensity. This index provides information
on the amount of  synchronization between the external forcing and the signal to control,
initially in a homoclinic chaotic regime. Performance has been evaluated for a random
uniform frequency distribution of the forcing signal centered in the middle value of 
the corresponding
Arnold tongue associated with the 1:1 phase locking, that is, centered at
$f_0=1.7$ $KHz$ (see Ref. \cite{alla01}). Thus, the frequency of the external modulation take values in
the interval $(f_0-f_s,f_0+f_s)$  provided that the 1:1 phase locking is guaranteed,
that is, for $f_s\le0.6$ $KHz$.


\begin{figure}[!h]
\begin{center}
\leavevmode \epsfxsize 80 mm \epsfbox{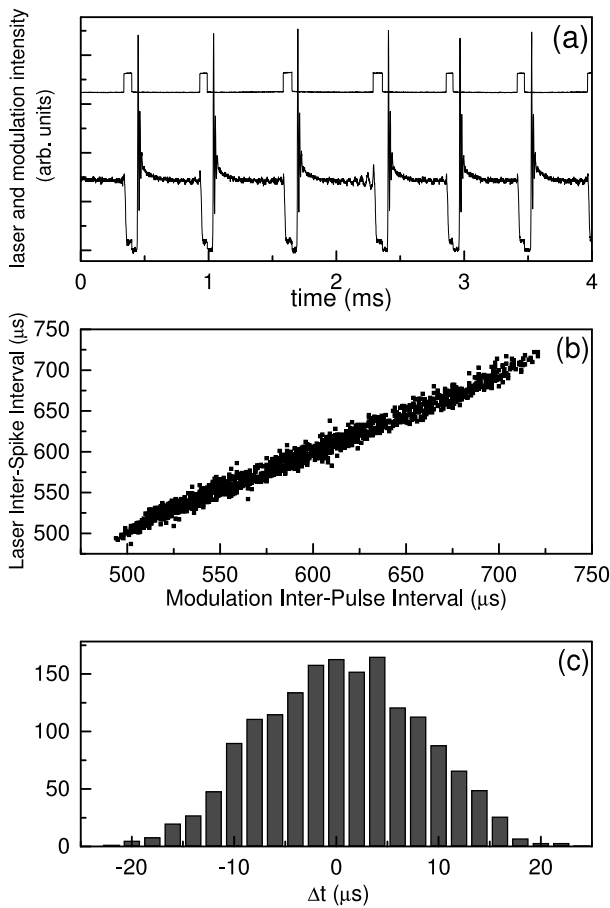}
\end{center}
\caption{(a) Experimental temporal evolution of  the laser intensity driven by an external pulsed signal.
(b) Inter-spike intervals of the laser intensity versus
 inter-pulse intervals of the forcing signal. This plot indicates synchronization between
 the two variables.
(c) Histogram of the time differences $(\Delta t)$ between inter-pulse intervals of the external signal 
and  inter-spike intervals of the laser intensity.}
\label{fig2}
\end{figure}

As shown in Fig. \ref{fig2}(a), the phase of the temporal series of the laser intensity follows 
almost perfectly the external signal, with the laser spikes 
adjusting  in each orbit  to the corresponding   external
signal. This is obtained because the laser parameters  are adjusted to
the  homoclinic chaos regime where the system crosses a region of large susceptibility
\cite{arec01}.
Figure \ref{fig2}(b) shows the inter-spike intervals of the laser intensity versus the 
inter-pulse intervals of the modulating signal. Synchronization is expressed by the straight line
of $45$ degrees. It is interesting to notice that synchronization 
does not depend on the frequency range of the external modulation. A better
estimation of the error in time between the inter-spike intervals of the laser and the timing
between  pulses of the external modulating signal is shown in Fig. \ref{fig2}(c), where an histogram 
of these differences in time
is plotted. Such histogram is fitted by  a zero-mean Gaussian distribution 
with a standard deviation of $7.8 \mu s$.
In this figure we also  appreciate that the difference between the
corresponding return periods of the external signal and the laser intensity is  always
below  a small value ($\Delta t\le25$ $\mu s$).


From this analysis, we conclude that it is possible to control the laser with pulses of very low
power. Thus, it is  possible to encode in the return periods of the laser intensity messages of 
binary, ternary, quaternary nature, etc...  The fact of encoding information
in the timing between spikes can be very useful in order to design a robust
communication system \cite{sush00} or to achieve a better understanding in the
communication phenomenon among cells in the central nervous system
\cite{schu01}. It is well known that the pattern of spike
times provides a large capacity for conveying information beyond that due
to the code commonly assumed by physiologists, the number of spikes fired
\cite{mack52}.


 \begin{figure}[!h]
 \begin{center}
 \leavevmode \epsfxsize 80 mm \epsfbox{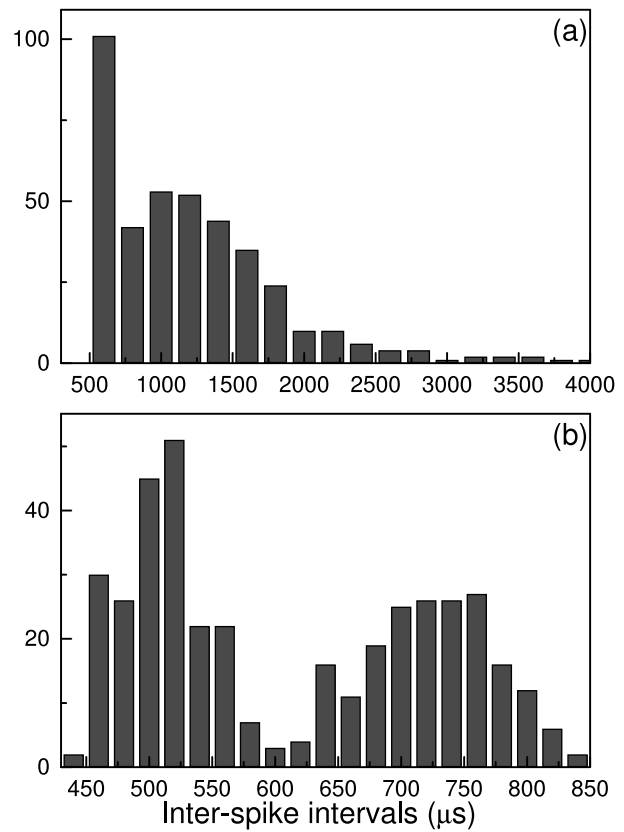}
 \end{center} 
 \caption{a) Histogram of
the frequencies of the free-running signal before applying any perturbation.
(b) Histogram  when some information has been encoded in the
signal. The information has been  considered as a uniform distribution of
bits '1' and bits '0'.}
 \label{fig3}
 \end{figure}

As an example of how to encode information in the timing between
spikes, we present the  simple case in which  a binary  message is
encoded in the return period of the spikes in the CO$_2$ laser intensity.
The  homoclinic chaotic behavior of the laser can be used to encode a desired
message in such a way that the bits ''1" and ''0" are easily identified as
timing between spikes less or greater than a threshold value.
The  encoding procedure is the following. First of all,
an  external pulsed signal that modulates the laser intensity is
synthesized with an inter-pulse frequency changed after  each pulse according to 
a uniform frequency distribution centered at the mean repetition
frequency, $f_0=1.7$ $KHz$. When the value of the frequency is larger
than  $f_0$,  a bit '1' is encoded in the
external signal;  when this value  is lower than $f_0$, a bit '0'  is encoded in the 
external signal.  Since the error in
time in the synchronization process between the external signal and the laser
intensity is below 25$\mu s$, in order to avoid errors around the mean
frequency,  we consider variations in the frequency in the interval
$(f_0-f_s,f_0-f_a)$ to encode the bit '0', and in the interval
$(f_0+f_a,f_0+f_s)$  to encode the bit '1'. It is obvious that for
$f_a\ge0.08$ $KHz$ no errors in the transmission are guaranteed, since this is
the variation in the frequency associated with a time variation of $25$
$\mu s$ $(\Delta f=\Delta t*f_0^2)$. Figure \ref{fig3}(a) shows the histogram
of the inter-spike intervals of the free-running signal before applying any
perturbation, and Figure \ref{fig3}(b) shows the same when some information
has been encoded in the own signal. The information has been  considered as a
uniform distribution of bits '1' and bits '0'. We can appreciate in these
histograms how a left shift in time (right shift in frequency) is produced
when the control is applied. This is easy to understand since the perturbation
tends to eliminate the oscillations of the laser intensity around the saddle
point, leading such variable to the rest state (no intensity). In this way, 
the orbit is shortened and the time difference between consecutive spikes is reduced.
 Notice also that the frequency range of the modulating 
signal to achieve  1:1 phase locking is centered around the most probable value of the
frequency corresponding to the homoclinic behavior of the laser intensity.


 \begin{figure}[!h]
 \begin{center}
 \leavevmode \epsfxsize 80 mm \epsfbox{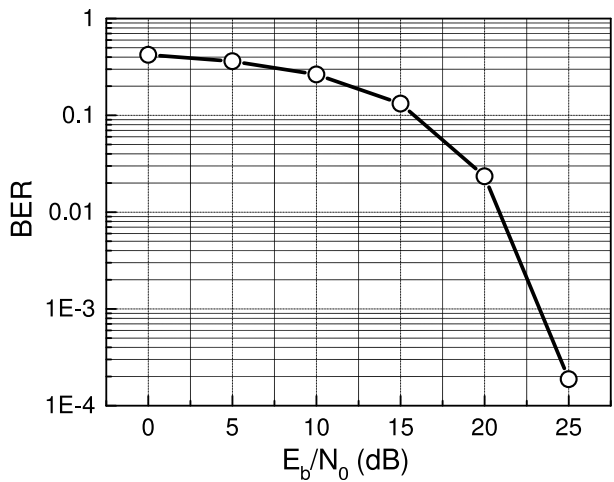}
 \end{center}
 \caption{BER vs. $E_b/N_0$ of the proposed communication system when additive
white Gaussian noise is contaminating the channel.}
 \label{fig4}
 \end{figure}

The information decoding  process  in the laser intensity can be
carried out by observing the instant when the laser intensity is higher than  a
threshold level. The fact that the spike amplitude is very high in
comparison with the signal elsewhere,
and the fact that the system can be controlled with small perturbations
of very low power, make it  suitable for
information transmission in a communication system. 
Since the information is contained entirely in the timing between spikes, 
channel distortions that affect the pulse shape will not significantly
influence the  information transmission.
From this point of view, we analyze how robust the system is
to the addition of  a zero-mean Additive White Gaussian Noise
(AWGN) in the communication channel. 
Notice that the channel noise is added  to the transmitted signal
as $s(t)+n(t)$ in the transmission channel and sent to the receiver.
Figure \ref{fig4} shows the Bit Error Rate (BER) versus $E_b/N_0$, where $E_b$ is the energy 
per bit, and $N_0/2$ represents the power spectral density of the channel
noise. We can see in this figure how the BER strongly decreases as
the $E_b/N_0$ relation increases. For values of $E_b/N_0\ge 25$ $dB$ the BER
is already below $10^{-3}$. Own method provides a BER consistently lower than
other recent proposals \cite{mura00}.

In conclusion, we have  shown experimentally how it is possible to control the inter-spike
intervals of a homoclinic chaotic system by applying small perturbations.
This fact allows to encode  information in the timing between spikes.
This can be useful in order to achieve a better understanding in the
communication phenomenon among cells in the central nervous system
as well as to  design a robust communication system.
In particular, we have shown a simple method to encode a binary message in a
homoclinic chaotic signal that can be used in communication applications.

We acknowledge  the support from the European Contract N$_0$. 
HPRN-CT-2000-00158.

\end{document}